\documentclass[runningheads]{llncs}
\usepackage{graphicx}

\usepackage[table]{xcolor}

\usepackage[num,german]{isodate}
\usepackage{datetime}
\usepackage[all]{background}
\SetBgContents{{\color{white} .}}
\SetBgPosition{15cm,1cm}
\SetBgAngle{0.0}
\SetBgScale{1.0}

\usepackage[T1]{fontenc}

\usepackage{comment}
\usepackage[backgroundcolor=yellow]{todonotes}
\usepackage{algorithm}
\usepackage[noend]{algpseudocode}
\usepackage{multirow}
\usepackage{xfrac}
\usepackage{dblfloatfix}    
\usepackage{float}

\RequirePackage[nomain,nonumberlist,nogroupskip]{glossaries}
\RequirePackage{xspace}
\RequirePackage{bm}
\RequirePackage{amsfonts}
\RequirePackage{textcomp}
\newglossary{symbols}{sym}{sbl}{List of Symbols}

\newglossary{hiddensymbols}{hidsym}{hidsbl}{} 

\newcommand{\sym}[4]{\newglossaryentry{#1}{type=symbols,
        sort=#2,
        name={\ensuremath{#3}\xspace},
        description={#4}}%
    \expandafter\newcommand\expandafter{\csname #1\endcsname}{\gls{#1}}%
}

\newcommand{\symHidden}[4]{\newglossaryentry{#1}{type=hiddensymbols,
        sort=#2,
        name={\ensuremath{#3}\xspace},
        description={#4}}%
    \expandafter\newcommand\expandafter{\csname #1\endcsname}{\gls{#1}}%
}

\newcommand{\symLower}[6]{\newglossaryentry{#1#2}{type=symbols,
        sort=#3,
        text={\ensuremath{#4}\xspace},
        name={\ensuremath{#4_{#5}}\xspace},
        description={#6}}%
    \expandafter\newcommand\expandafter{\csname #1\endcsname}[1]{\ensuremath{\gls{#1#2}_{##1}}}%
    \expandafter\newcommand\expandafter{\csname #1#2\endcsname}{\ensuremath{\gls{#1#2}_{#5}}}%
}

\newcommand{\symLowerComma}[6]{\newglossaryentry{#1#2}{type=symbols,
        sort=#1#2,
        text={\ensuremath{#3}\xspace},
        name={\ensuremath{#3_{#4,#5}}\xspace},
        description={#6}}%
    \expandafter\newcommand\expandafter{\csname #1\endcsname}[1]{\gls{#1#2}_{#4,##1}}%
    \expandafter\newcommand\expandafter{\csname #1#2\endcsname}{\gls{#1#2}_{#4,#5}}%
}

\newcommand{\symFun}[6]{\newglossaryentry{#1#2}{type=symbols,
        sort=#3,
        text={\ensuremath{#4}\xspace},
        name={\ensuremath{#4\left({#5}\right)}\xspace},
        description={#6}}%
        \expandafter\newcommand\expandafter{\csname #1\endcsname}[1]{\gls{#1#2}\!\left({##1}\right)}%
        \expandafter\newcommand\expandafter{\csname #1#2\endcsname}{\gls{#1#2}\left({#5}\right)}%
}

\symLower{loss}{AB}{LAB}{\Lambda}{A,B}{Loss between node $A$ and $B$}
\sym{bound}		{l}{\hat{\lambda}}{The current path loss bound}
\sym{boundbest}		{lbest}{\hat{\lambda}_{\text{best}}}{The path loss bound corresponding to the maximum depth}
\sym{depth}		{d}{\delta}{The current depth}
\sym{margin}		{D}{\Delta}{Path loss margin}
\sym{depthbest}		{dbest}{\delta_{\text{best}}}{The maximum depth found}
\sym{selbest}		{Sbest}{S_{\text{best}}}{The set of the best node and bound selections}
\sym{nodebool}		{x}{\bm{x}}{Vector to determine which nodes to select}
\symLower{boundedneighbors}{ul}{Nul}{\mathcal{N}}{u,\hat{\lambda}}{Set of neighbors of $u$ with a loss smaller than or equal to $\hat{\lambda}$}
\symFun{dreq}{d}{kd}{\kappa}{\delta}{Minimum required number of nodes per level}

\usetikzlibrary{chains}

\begin{document}

\title{Constructing Customized Multi-Hop Topologies\\ in Dense Wireless Network Testbeds}

\titlerunning{Constructing Customized Multi-Hop Topologies}

\author{Florian Kauer and Volker Turau}

\institute{Institute of Telematics\\Hamburg University of Technology, Hamburg, Germany,
\email{\{florian.kauer,turau\}@tuhh.de}}
\maketitle              

\begin{abstract}
  Testbeds are a key element in the evaluation of
  wireless multi-hop networks. In order to relieve researchers from
  the hassle of deploying their own testbeds, remotely controllable
  testbeds, such as the FIT/IoT-LAB, are built. However, while the
  IoT-LAB has a high number of
  nodes, they are deployed in constraint areas. This, together with the complex nature
  of radio propagation, makes an ad-hoc construction of multi-hop
  topologies with a high number of hops difficult. This
  work presents a strategic approach to solve this problem and
  proposes algorithms to generate topologies with desired
  properties. The implementation is evaluated for the IoT-LAB testbeds
  and is provided as open-source software. The results
  show that preset topologies of various types can be built even in
  dense wireless testbeds.
\end{abstract}

\section{Introduction}
\begin{figure}
\centering
\begin{tikzpicture}
  \node{\includegraphics[scale=0.5]{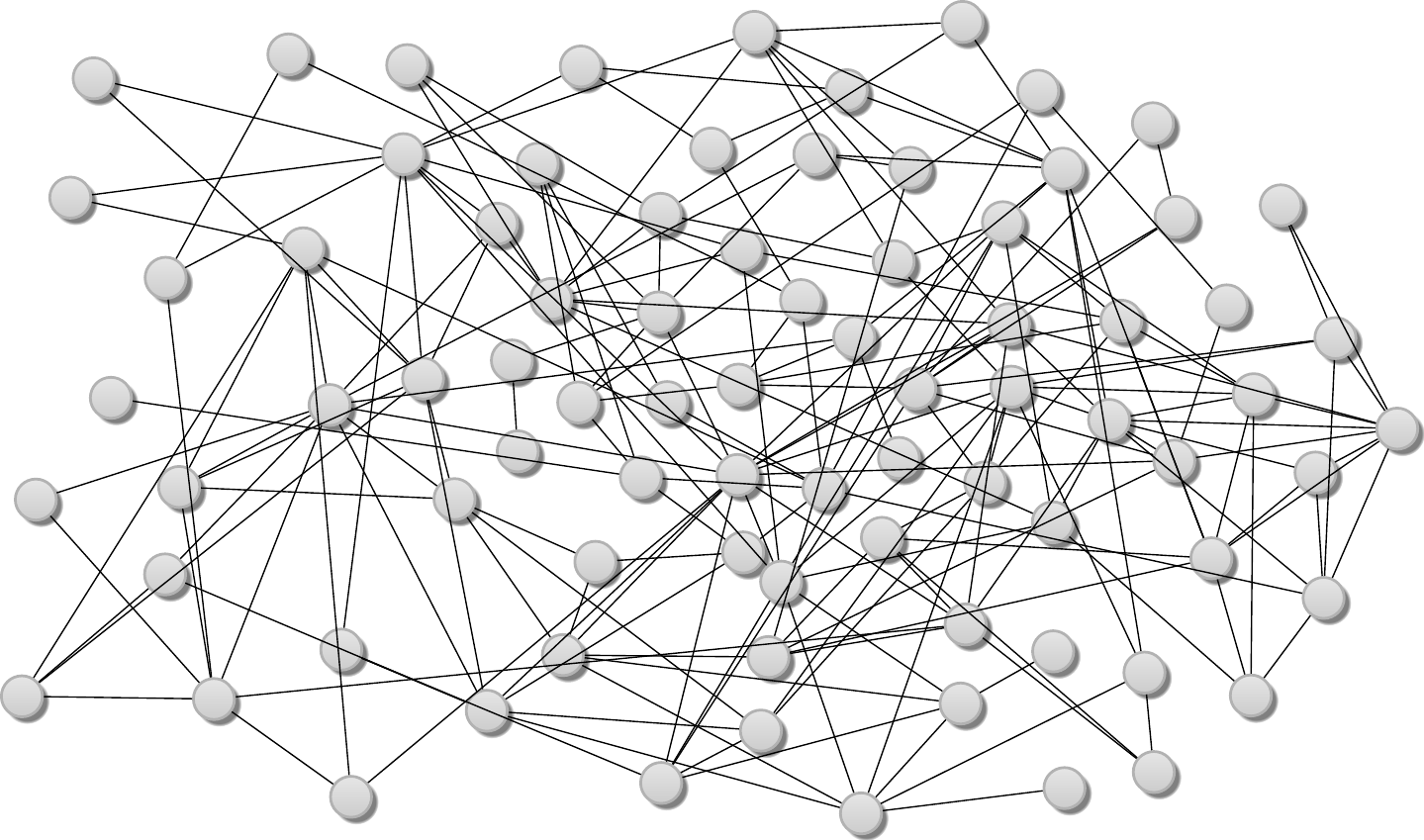}};
  \node[xshift=4cm,yshift=0.28cm]{\Large $\rightarrow$};
  \node[xshift=6.5cm,yshift=0.28cm]{\includegraphics[scale=0.5]{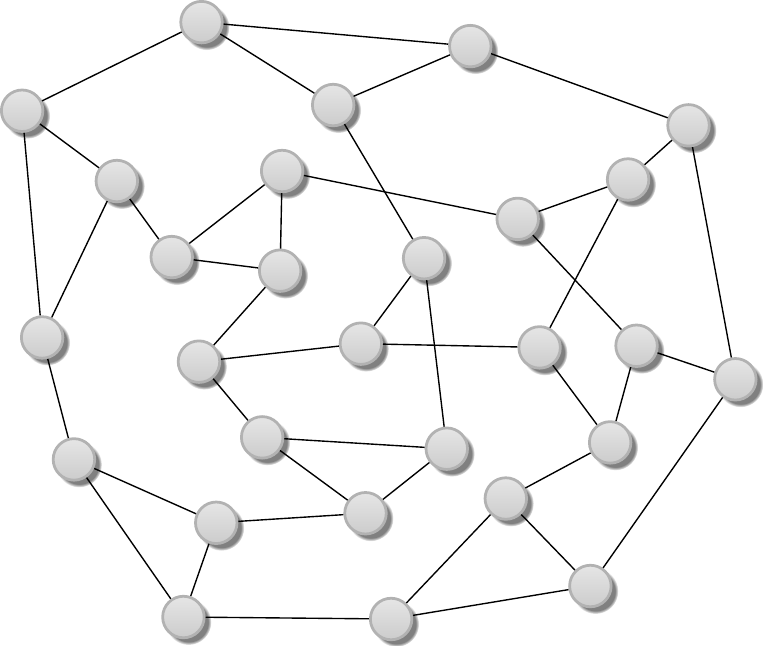}};
\end{tikzpicture}
      \caption{Construction of an induced subgraph with uniform density.\label{fig:degredlille}}
    \vspace{-1cm}
\end{figure}

The use of testbeds consisting of actual wireless hardware is of major importance for development and evaluation of algorithms and protocols for wireless networks. While analytical considerations and simulations come with less initial investment, they can only partly reproduce the real world. The main reason is that propagation of radio waves is highly complex and even the most complex models can only cover parts of the actual mechanisms and are computationally expensive.
However, the deployment of wireless hardware comes with a high effort, especially when targeting large-scale multi-hop networks to be used in applications such as industrial plants \cite{odonovan_ginseng_2013,pfahl_holistic_2014}. Instead of setting up a new testbed it is therefore often advisable to resort to existing testbeds that provide convenient remote control interfaces.
One popular example is the FIT/IoT-LAB \cite{iotlab} that consists of multiple wireless network deployments ranging from 41 to 928 nodes of different types.
These numbers are sufficient for many experiments. However, most deployments of the IoT-LAB span a comparatively small area where every transceiver can reach every other node in one hop, which does not allow evaluating complex routing algorithms, such as RPL~\cite{rfc6550}, or data link layers such as IEEE 802.15.4 DSME~\cite{towards_openDSME}. The evaluation of such protocols requires dedicated topologies with specific properties. Furthermore, during debugging, very specific topologies can be very helpful.
One obvious solution is to reduce transmission power and receiver sensitivity of the transceivers or to select a subset of nodes to generate more sparsely connected networks. By this approach it is also possible to setup specific conditions to induce certain phenomena. This is especially interesting to provoke and debug weaknesses of protocols. Relevant topology properties include density -- to regulate channel utilization and spatial reuse --, a high number of hops or the (non-)existence of weak or asymmetric links. Building long chains is interesting to model tunnels \cite{ceriotti_is_2011}, grids for potato fields \cite{langendoen_murphy_2006} or tree-like structures for large-scale data collection \cite{mao_citysee:_2012}.
The main contribution of this paper is an approach to construct topologies in  existing sensor network deployments satisfying given conditions for evaluating multi-hop protocols based on link quality measurements. The focus is on the construction of subgraphs with uniform density as illustrated in Fig.~\ref{fig:degredlille} as well as on tree-like topologies with specified depth and breadth. The implementation is published as open-source software \cite{topologies_implementation} to enable other researchers to generate suitable topologies for their experiments in the testbed of their choice. We present applications of our method for deployments in the FIT/IoT-LAB.

\subsection{Related Work}

Building large-scale wireless sensor networks, especially outdoors, is a challenging task. Successful examples include 
the Trio Testbed with 557 nodes in an area of $50000\,\text{m}^2$ \cite{dutta_trio:_2006} and CitySee \cite{mao_citysee:_2012} with 1196 nodes that build a multi-hop topology with up to 20 hops.
Valuable overviews over the numerous problems that have to be solved are given in \cite{barrenetxea_hitchhikers_2008} and \cite{langendoen_murphy_2006}.
Therefore, multiple initiatives are made to build generic testbeds that can be controlled remotely by researches all around the world. These includes WISEBED \cite{chatzigiannakis_wisebed}, the MoteLab \cite{werner-allen_motelab:_2005} and the FIT/IoT-LAB \cite{iotlab}, providing convenient remote interfaces for software updates and debugging.
Several publications cover channel characterization in the IoT-LAB \cite{bildea_link_2013,watteyne_lessons_2015}, including work on the applicability of RSSI measurements for localization \cite{heurtefeux_is_2012} and with the focus on the repeatability of experiments \cite{papadopoulos_importance_2016,papadopoulos_thorough_2017}. The latter covers many aspects relevant for this paper, including transmission power selection for controlling the density and the selection of quality radio links. For this, a strategic approach is important, because the topology has a large impact on the performance, for example on delivery ratio and energy consumption, as discussed in \cite{ducrocq_impact_2014}.
Outside the IoT-LAB, channel characterizations were conducted for example in complex factory environments \cite{tang_channel_2007} and by evaluating the influence of antenna, mutual alignment and distance on the transmission between wireless sensor nodes \cite{myers_experimental_2007}. In the same publication, the influence of transmission power adjustment is discussed to minimize interferences.
The adjustment of the transmission power for topology control is a broadly studied topic \cite{ramanathan_topology_2000,li_survey_2013,santi_topology_2005}. The latter also covers homogeneous transmission power adjustment for topology control and is therefore similar to the approach in this paper, though the objective is different.

\section{Emulating Channel Conditions}
One cornerstone of the proposed approach is the possibility to emulate different channel conditions by manipulating the transmission power and sensitivity of the transceivers. The primary node type in the IoT-LAB is the M3 Open Node. It consists of an ARM Cortex M3, an Atmel AT86RF231~\cite{at86rf231}, a $2.4\,\text{GHz}$ chip antenna and several other peripherals. The transceiver can be configured with an output power from $-17\,\text{dBm}$ to $3\,\text{dBm}$ and a reception sensitivity from $-48\,\text{dBm}$ to $-101\,\text{dBm}$.
The overall reduction of signal power between two nodes $a$ and $b$ is denoted as $\loss{a,b}$ in the following. It depends on various losses and gains, including potential losses between the transceivers and the antennas, the gain of the antennas, which highly depends on the mutual alignment of the transceivers, and the path loss, which might also include fading effects, such as multi-path propagation. For example, given a link with $\loss{a,b}=63\,\text{dB}$, a communication will be possible for a sensitivity setting of $-66\,\text{dBm}$ if the transmission power is chosen to be larger than $-3\,\text{dBm}$. If the sensitivity is not reduced, reliable communication is even possible for the lowest transmission power of $-17\,\text{dBm}$.
\vspace{-1mm}
\section{Topology Generation Procedure}
\vspace{-1mm}
In this section, the proposed procedure for generating multi-hop topologies is explained and exemplarily conducted for the Saclay testbed, more specifically the 12 M3 Open Nodes aligned in a 3x4 grid in the so called Digiteo 2 room.
Compared to the other testbeds of the IoT-LAB it is relatively small and compact, so it allows for better traceability of the results in the following. Full results for the other testbeds are given in Sect.~\ref{sect:comparison}. The general procedure is as follows:
\vspace{-2mm}
\begin{enumerate}
  \item Measure $\loss{a,b}$ between every pair of nodes.
  \item Estimate the neighborhood graphs depending on the transmission power and sensitivity setting.
  \item By means of these graphs, construct a topology by selecting a subset of the nodes and appropriate settings.
  \item Verify this selection in the real testbed. 
\end{enumerate}
\begin{figure*}[tbh]
\centering
  \begin{tikzpicture}
    \node (a) {\includegraphics{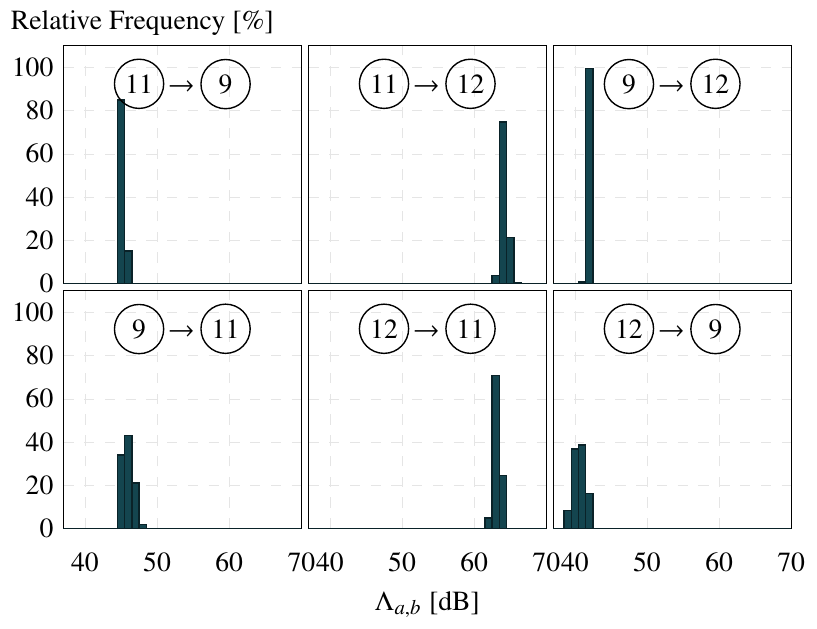}};
    \node[anchor=west,xshift=-0.2cm] at (a.east) {\includegraphics{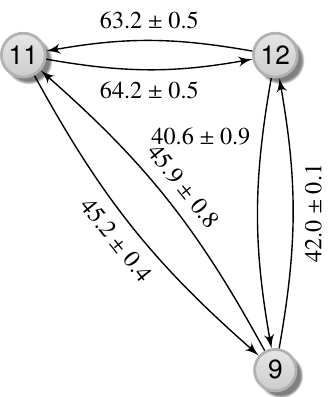}};
  \end{tikzpicture}%
  \vspace{-0.3cm}
\caption{The distribution of the pairwise RSSI measurements for three nodes. The right part shows the associated graph of the nodes together with the resulting mean and standard deviation of the measurements.\label{fig:singlelinks}}
 \vspace{-0.4cm}%
\end{figure*} %
\subsection{Pairwise $\loss{a,b}$ Measurements}
The RSSI is used in the following to estimate $\loss{a,b}$ between each pair of nodes in the deployments of the IoT-LAB. Please be advised that wireless conditions change rapidly and are hardly reproducible, especially when the conditions change, such as a window or a door that is opened. Therefore, recent measurements will get better results, but still the topology generation algorithm itself has to take these fluctuations into account. The measurement is conducted as follows. Every node repeatedly sends out packets with full transmission power. To reduce the probability of collisions, random intervals between the transmissions are chosen and CSMA/CA is used according to the IEEE~802.15.4 standard. Every node in the neighborhood where the signal is strong enough to be received, measures the RSSI of the packet. By subtracting the RSSI from the constant and known transmission power, we get an estimate for $\loss{a,b}$ between the respective nodes. This is repeated so that finally every node has sent at least 250 packets. It is important to measure on the same frequency channel as the one used in the final experiment. When multiple channels will be used such as in DSME \cite{towards_openDSME}, all relevant channels have to be taken into account.
\begin{figure}[bt]
\centering
    \begin{minipage}{.45\columnwidth}
\centering\includegraphics{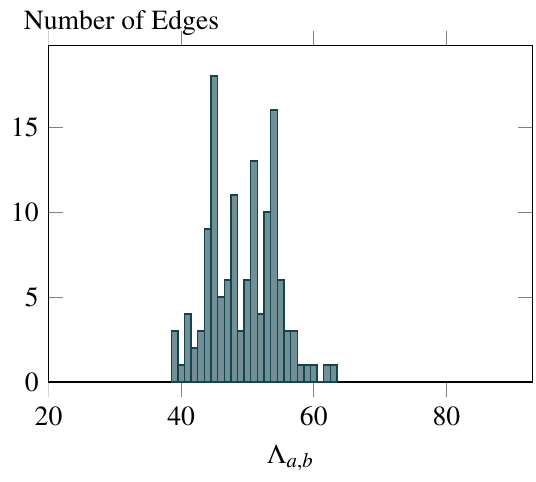}
  \vspace{-0.75cm}
  \caption{The distribution of the measured $\loss{a,b}$ for the Saclay testbed.\label{fig:linkhist}}
  \vspace{0.75cm}
    \end{minipage}\hspace{10mm}\begin{minipage}{.45\columnwidth}
\centering\includegraphics{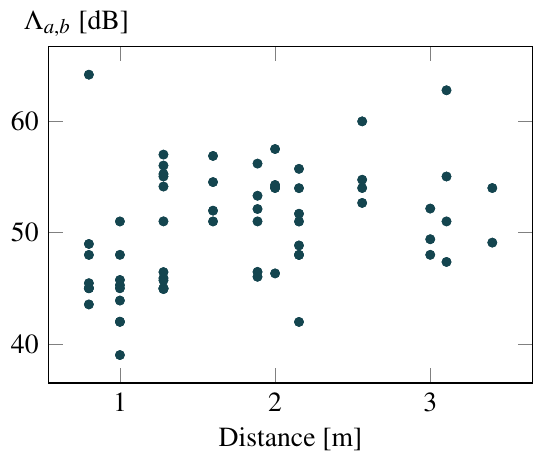}%
  \vspace{-0.3cm}
\caption{Measured $\loss{a,b}$ in relation to the euclidean distance between the nodes showing only a slight correlation (correlation coefficient $0.38$).\label{fig:distance}}%
    \end{minipage}
  \vspace{-0.5cm}
\end{figure} %
Fig.~\ref{fig:singlelinks} exemplarily shows the results for three nodes in the Saclay deployment. The left part of the figure depicts the distribution of the measured value for every pair of nodes, while the right part shows the mean and standard deviation of these measurements on the graph. Fig.~\ref{fig:linkhist} shows a histogram over all links between the 12 M3OpenNodes and in Fig.~\ref{fig:distance} the losses over the distance between the respective nodes are plotted, showing only a slight correlation with a correlation coefficient of $0.38$. This is also the reason why generating a topology by picking seemingly fitting nodes in the map of physical locations rarely works well.

\subsection{Neighborhood Graphs}

\begin{figure*}[b]
  \vspace{-0.3cm}
\centering%
\includegraphics{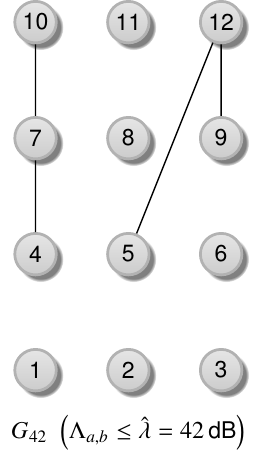}%
  \hspace{0.48cm}%
\includegraphics{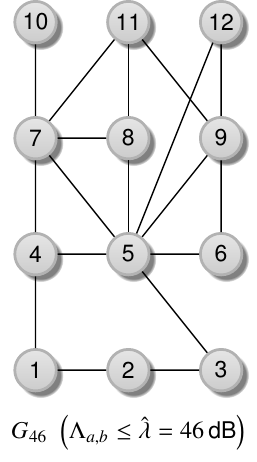}%
  \hspace{0.48cm}%
\includegraphics{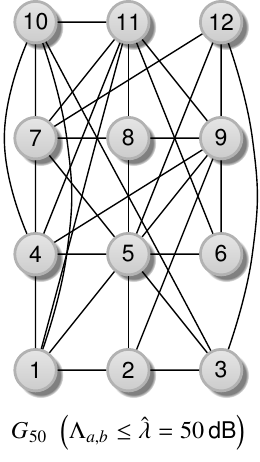}%
  \hspace{0.48cm}%
\includegraphics{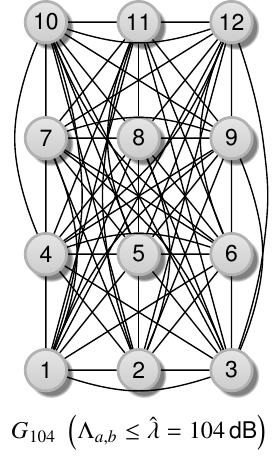}%
  \caption{Selected graphs from $\bm{G}$ for the Saclay testbed.\label{fig:saclaytopo}}
\end{figure*} %

Based on the data from the last section, the existence of usable bidirectional links for various link budgets can be estimated. An edge is added to the neighborhood graph if the measured path loss between two nodes for both directions is smaller than the given link budget. This gives a family $\bm{G}$ of graphs $G_{{\bound}}=\left(V,E_{\bound}\right)$, with $\bound_{min} \leq \bound \leq \bound_{max}$ and
$E_{\bound} = \left\{(a,b)\,|\,\loss{a,b} \leq \bound \wedge \loss{b,a} \leq \bound\right\}.$
Here, $\bound_{min} = 31\,\text{dB}$ is the smallest and $\bound_{max} = 104\,\text{dB}$ the largest bound that is possible with the AT86RF231.  
Fig.~\ref{fig:saclaytopo} shows the resulting graphs $G_{42}$, $G_{46}$, $G_{50}$ and $G_{104}$ for the Saclay testbed. Obviously, the number of edges increases with $\bound$. $G_{104}$ is fully meshed, every node can reach every other node in a single hop. It is also apparent that there exist good links of long euclidean distance (e.g. $5 \leftrightarrow 12$), but also bad links with small euclidean distance (e.g. $11 \leftrightarrow 12$). Reasons for this include nonuniform antenna patterns, obstacles and reflections. 

\begin{figure}[h]%
\centering\includegraphics{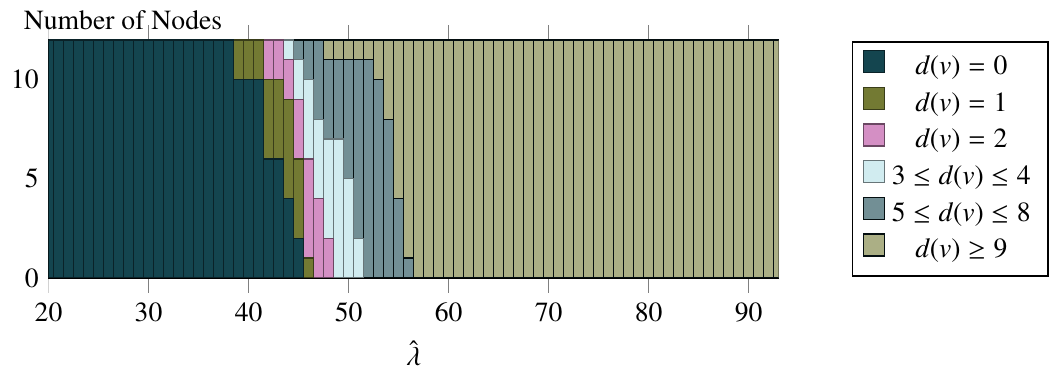}%
\vspace{-0.5cm}
\caption{The distribution of the node degree over the bound $\bound$.\label{fig:nodedeg}}%
\vspace{-0.5cm}
\end{figure} %

The analysis of these graphs will finally help to decide which transmission power and which sensitivity should be chosen for a suitable topology for evaluation of multi-hop topologies. One approach to this task is the analysis of the degrees $d(v)$ of the nodes in the network. In Fig.~\ref{fig:nodedeg}, the distribution of the node degrees is shown for each $\bound$ for the Saclay testbed. Beyond $\bound = 56$, all nodes have at least 9 neighbors, while below $\bound = 39$, no edges exist. The intermediate section is the most interesting for our application. For example, in $G_{46}$, that is also shown in Fig.~\ref{fig:saclaytopo}, one node has only one neighbor, five have 2 neighbors, four have degree 3 or 4, and for two it is between 5 and 8.

\subsection{Construct a Topology with Given Density}
The task of this step is to decide for a subset of the nodes that form a topology suitable for the respective requirements at hand and a bound $\bound$, corresponding to the respective transmission power and sensitivity settings. While different settings per node may be feasible for some scenarios, we choose a homogeneous $\bound$ to simplify the final realization and to avoid unexpected effects from heterogeneous transmission powers. As written in the introduction, the requirements for this selection can be very diverse. While very dense graphs with few hops can be realized easily by selecting as many nodes as possible and using high transmission power and sensitivity, the construction of the longest chain of nodes requires solving a well known NP-complete problem, referred to as \emph{induced path} \cite{garey_computers_1979}.

At first we consider  topologies with constant node degree $c$, i.e., every node has $c$ neighbors. This problem can be formulated and solved by integer linear programming (ILP). Let $\nodebool \in \{0,1\}^{\left|V\right|}$ such that $\nodebool(u) = 1$ if node $u$ is selected and $\nodebool(u) = 0$ if not. The set of neighbors of node $u$ with a loss at most $\bound$ is given by
    $\boundedneighbors{u,\bound} = \left\{v \,\middle|\, (u,v) \in E_{\bound}\right\}.$
For a given $\bound$, the ILP can be formulated as:
\begin{align*}
  \begin{aligned}
    & \underset{\nodebool}{\text{maximize}}
    & & \sum_{u\,\in\,V} \nodebool(u) \\
      & \text{subject to}
      & & c \cdot \nodebool(u) \leq \sum_{v\,\in\,\boundedneighbors{u,\bound}} \nodebool(v) \leq c + m \cdot \left(1-\nodebool(u)\right), \forall u \in V.
  \end{aligned}
\end{align*}

\begin{figure}[tb]
\begin{minipage}{.5\columnwidth}
\centering
	\includegraphics{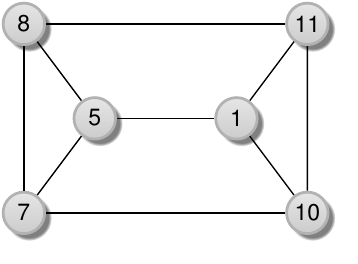}
    \end{minipage}\hspace{0mm}\begin{minipage}{.5\columnwidth}
\centering
	\includegraphics{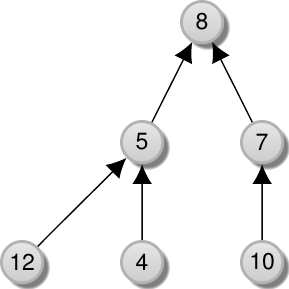}
\end{minipage}
\begin{minipage}{.48\columnwidth}
    \caption{Topology generated by the first technique with $\bound=49\,\text{dB}$ and $c=3$. The positions do not represent the physical locations (cf. Fig.~\ref{fig:saclaytopo}).\label{fig:saclaydeg}}
\end{minipage}\hspace{0.04\columnwidth}\begin{minipage}{.48\columnwidth}
  \caption{RPL routing tree for a transmission power of $-17\,\text{dBm}$ and a sensitivity of $-66\,\text{dBm}$. Topology generated by the second technique.\label{fig:rpl}}
\end{minipage}
\end{figure}

Here, $m=\max_{w\in V}\left(\left|\boundedneighbors{w,\bound}\right|\right)$, so for $\nodebool(v)=0$ the condition reduces to
$0 \leq \sum_{v\,\in\,\boundedneighbors{u,\bound}} \nodebool(v) \leq c + m$
and therefore always holds, while for $\nodebool(u)=1$, it is equivalent to $\sum_{v\,\in\,\boundedneighbors{u,\bound}} = c$. This can for example be solved by using the COIN-OR Cbc \cite{lougee-heimer_common_2003} solver that is interfaced with the Python PuLP frontend in our implementation. By iterating over the $\bound$, we get a set of (not necessarily connected) subgraphs and can then select, for example, the largest connected component from these. Fig.~\ref{fig:saclaydeg} shows a topology for the Saclay testbed that is generated by this procedure. Also Fig.~\ref{fig:degredlille} is the result of applying this technique to the largest connected component for $\bound = 47\,\text{dB}$ and $c=3$ in the Lille testbed.

\subsection{Construct a Tree Topology}
For many evaluations, especially for analyzing tree routing techniques such as RPL, a tree topology with a large depth is useful. The following properties allow for a versatile, yet easy to compute, construction of such tree topologies.
\begin{enumerate}
\item The graph is connected.
\item A node $v_0$ is designated as root, e.g., to serve as a RPL DODAG root.
\item The number of nodes that are reachable from $v_0$ over exactly $\depth$ hops, referred to as breadth in this paper, is at least $\dreq{\depth}$, being a predefined function. For example, with $\dreq{\depth} = \depth+1$, the number of nodes per level increases, while $\dreq{\depth} = 1$ also allows for linear topologies.
\item There exist no links with a $\loss{a,b}\le \bound+\margin$ that
  would change the topology if the conditions change slightly;
  $\margin$ is a margin to account for fluctuations.
\end{enumerate}

\algdef{SE}[DOWHILE]{Do}{doWhile}{\algorithmicdo}[1]{\algorithmicwhile\ #1}%

\begin{algorithm}[tb]
    \caption{Algorithm to construct a leveled subgraph for a given function $\kappa$}
  \label{algo:topo}
  \begin{algorithmic}[1]
    \Procedure {MonitoredBFS}{$V,E,v_0,\bound,\margin$}
      \State $V_{sub}\left(0\right) \leftarrow \left\{v_0\right\}$, $\depth \leftarrow 0$
      \Do
	\State $V_{sub}(\depth+1) \leftarrow\left\{\right\}$
	\ForAll{$u \in V_{sub}(\depth)$}
      \ForAll{$v \in \boundedneighbors{u,\bound}$}
	    \If{$v \notin \bigcup_{i = 0}^{\depth+1} V_{sub}(i)$}
        \If{$\boundedneighbors{v,\bound+\margin} \cap \bigcup_{i = 0}^{\depth-1} V_{sub}(i) = \emptyset$}\label{algline:weak}
	      \State $V_{sub}(\depth+1) \leftarrow V_{sub}(\depth+1) \cup \{v\}$
	    \EndIf
	    \EndIf
	  \EndFor
	\EndFor
	\State $\depth \leftarrow \depth + 1$
      \doWhile{$\left|V_{sub}(\depth)\right| \geq \dreq{\depth}$} \label{algline:growing}
      \State $V_{sub}(\depth) \leftarrow \{\}$ \Comment{remove partially filled layer}
      \State \Return $\left(\depth-1, V_{sub}\right)$
    \EndProcedure
  \end{algorithmic}
\end{algorithm}

Algorithm~\ref{algo:topo} generates a subsets of nodes for given $\bound$ and $v_0$ that lead to the largest number $\depthbest$ of hops towards $v_0$. In the algorithm, $V_{sub}: \mathbb{N}_0 \to \mathcal{P}(V)$ associates a depth value with the selected nodes of this depth.

Procedure \textproc{MonitoredBFS} is basically a breadth-first search
starting from $v_0$ and thus guarantees requirement 1 and 2. The
condition in line \ref{algline:growing} monitors the number of nodes
per level and aborts the search when it can not be continued without
violating requirement 3. Finally, line \ref{algline:weak} ensures
requirement 4 by excluding nodes that would have been visited earlier
when $\bound$ would have been selected slightly larger. It has to be
noted that this algorithm does not necessarily find \emph{the} best
possible topology, because it might be possible to generate subgraphs
with a larger depth by removing nodes a priori. Though, this again
would lead to the longest induced path problem.  \textproc{MonitoredBFS} is called for all values of $\bound$ and all nodes $v_0$ to get a set of subgraphs where we can select, for example,
one of the subgraphs with the largest depth.

\subsubsection{Node Reduction}
Some applications require a large number of nodes, but usually it is advisable to reduce the size of the network to ease the analysis of a particular phenomenon as long as the behavior is unchanged. It should also not be forgotten that the FIT/IoT-LAB is shared with other researchers, so less nodes means less hindrance for others. Therefore, this section presents an optional procedure to reduce the number of nodes of the previously found subgraph, while still maintaining the requirements. It basically strips away all nodes that are not on a path to a higher depth with the additional constraint of maintaining $\kappa\left(\depth\right)$ nodes for depth $\depth$. For a given generated layered subset of nodes $V_{sub}$ with associated $\depth$ and $\bound$, this can again be specified as ILP where we want to
\begin{align*}
  \begin{aligned}
    & \underset{\nodebool}{\text{minimize}}
    & & \sum_{u\,\in\,V} \nodebool(u) \\
      & \text{subject to}
      & & \sum_{u\,\in\,V_{sub}(i)} \nodebool(u) \geq \dreq{i} \quad \forall 1 \leq i \leq \depth\\
      &&&  \forall 1 \leq i \leq \depth, \forall u \in V_{sub}(i):\\
      & & & \nodebool(u) \leq \sum_{v\,\in\,V_{sub}(i-1)\,\cap\,\boundedneighbors{u,\bound}} \nodebool(v).
  \end{aligned}
\end{align*}

\subsection{Verification}
\label{sect:verification}
The output of the algorithm is based on measurements of the pairwise RSSI values which are fluctuating due to changes in the environment. Therefore, current measurements are necessary if a high accuracy is required and it is important to verify that the found topology in fact fulfills the given requirements before starting the actual experiment. 

For this, we use the RPL TSCH example for the IoT-LAB that is included in the IoT-LAB Contiki fork. A topology found for the given measurements of the Saclay testbed and $\dreq{\depth} = \depth+1$ consists of the nodes $\left\{4,5,7,8,10,12\right\}$ with root $v_0 = 8$ for the bound $\bound = 46\,\text{dB}$. Due to the constraint size of the testbed, it has depth $\depthbest = 2$. It can for example be achieved by choosing a transmission power of $-17\,\text{dBm}$ and a sensitivity of $-63\,\text{dBm}$, since%
$-17\,\text{dBm}-\left(-63\,\text{dBm}\right) = 46\,\text{dB}$.

It is, however, important to consider the transition region between perfect and no reception. Thus, it may be necessary to increase the transmission power slightly or to improve the sensitivity. For this evaluation, the sensitivity was improved to $-66\,\text{dBm}$. Finally, we get the routing tree by requesting the RPL parent for each node. This results in the tree shown in Fig.~\ref{fig:rpl} fulfilling the given requirements.

\section{Testbed Comparison}
\label{sect:comparison}

\begin{figure}[p]
\centering
	\includegraphics{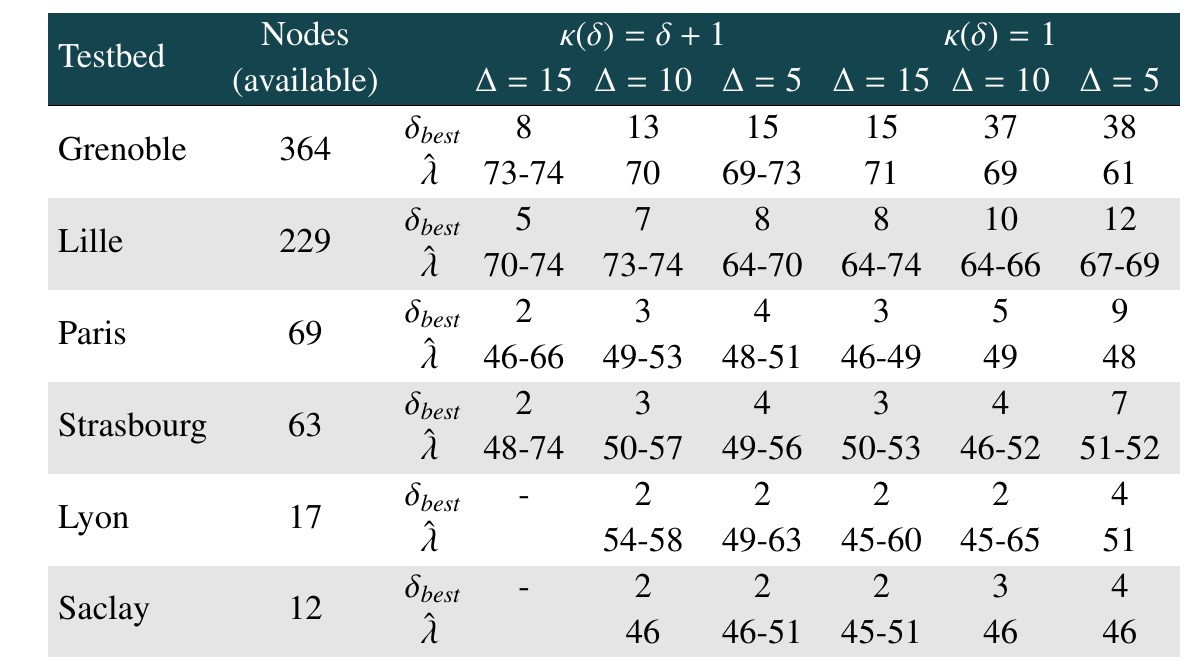}
    \caption{Maximal achievable depth $\depth$ and the associated $\bound$ range in dB for the different testbeds of the FIT/IoT-LAB. In Lyon and Saclay, no topologies with at least two hops are possible for $\margin = 15$ and $\kappa\left(\delta\right)=\delta+1$.\label{fig:restable}}
\end{figure} %
\begin{figure}[p]
      \vspace{-0.2cm}
    \begin{minipage}{.35\columnwidth}
\centering
    \includegraphics[scale=0.57]{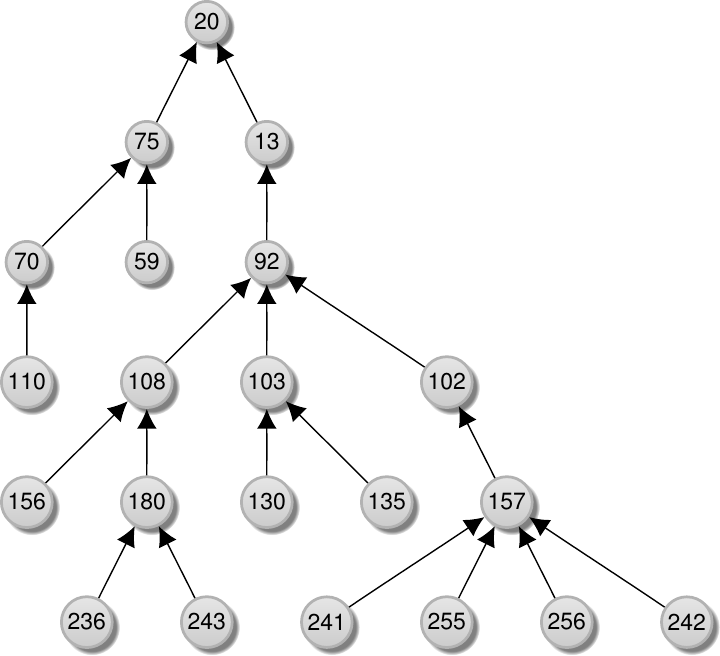}
    \end{minipage}\hspace{0mm}\begin{minipage}{.55\columnwidth}
    \includegraphics[scale=0.57]{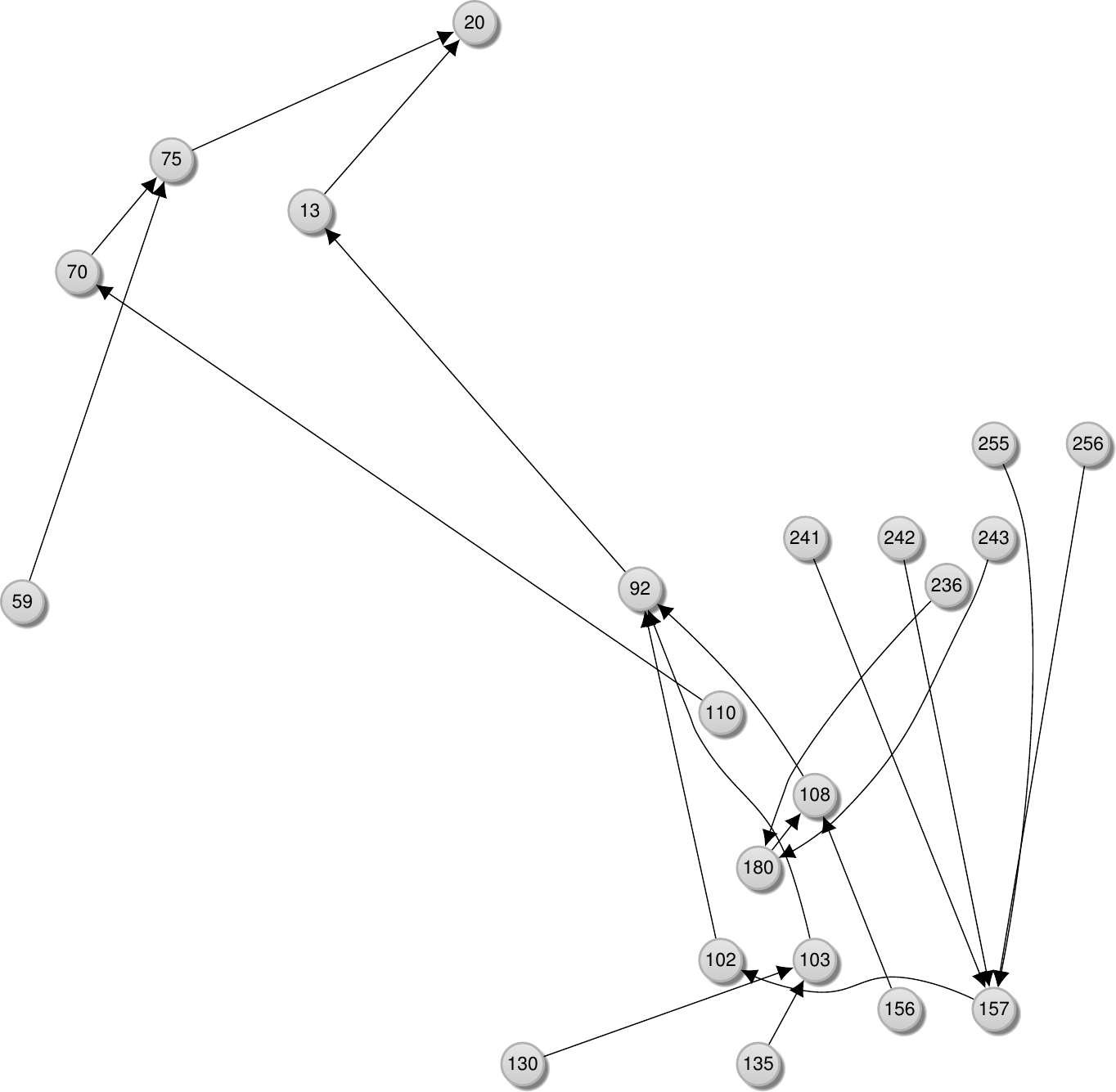}
    \end{minipage}
    \begin{minipage}{.35\columnwidth}
  \caption{A RPL routing tree with depth 5 at the Lille site for a transmission power of $3\,\text{dBm}$ and a sensitivity of $-81\,\text{dBm}$.\label{fig:rpl-lille}}
    \end{minipage}\hspace{3mm}\begin{minipage}{.625\columnwidth}
      \caption{The tree of Fig.~\ref{fig:rpl-lille} with the physical positions of the nodes.\label{fig:rpl-lille-pos}\vspace{3.6em}}
    \end{minipage}
    \vspace{-0.3cm}
\end{figure} %

After focusing on the Saclay testbed we consider other testbeds in this section. Fig.~\ref{fig:restable} shows the result for applying the second approach to all testbeds. The larger testbeds, Grenoble and Lille are in fact the only  that are not fully meshed at full transmission power and sensitivity. With the previously stated requirements and $\margin=15$, the maximum achievable depth for $\dreq{\depth} = \depth+1$ according to the measurements is $\depthbest=8$ when setting $\bound$ to $74\,\text{dB}$ in Grenoble. With a reduced $\margin=5$, the depth increases to 15, but the topology is less robust. Also, with $\dreq{\depth}=1$ more hops can be achieved. Finally, Fig.~\ref{fig:rpl-lille} shows a resulting RPL routing tree in the Lille testbed and Fig.~\ref{fig:rpl-lille-pos} depicts the physical positions of the nodes in this experiment with some links that would not be obvious based on the locations, demonstrating the benefit of the presented approach.

\vspace{-0.2cm}
\section{Conclusion}
\label{sect:conclusion}
The paper proposes an approach to generate multi-hop topologies in dense wireless network testbeds. The well-known fact that in common wireless sensor network settings, the received signal strength only correlates slightly with the distance can be verified for the FIT/IoT-LAB testbed. This makes it difficult to handpick reasonable nodes and settings for executing experiments. Therefore, channel condition measurements are conducted to estimate neighborhood graphs depending on the transmission power and sensitivity settings. As expected, the density of the resulting graph decreases with lower transmission power and reduced sensitivity. 

These measurements form the starting point for a constructive algorithm to generate tree topologies with a custom minimum number of nodes per depth as well as an ILP for reducing the number of nodes afterwards. Finally, it is shown that by the proposed approach the number of hops can be increased significantly in contrast to the default settings, where - except for Grenoble and Lille - only single-hop topologies are possible.

The published toolset \cite{topologies_implementation} is a convenient alternative for the conventional map based node selection in the FIT/IoT-LAB to be used by other researchers. Future work includes the development of alternative approaches for selecting nodes, for example to generate topologies appropriate for peer-to-peer experiments. Using a strategic approach for experiment setup as proposed in this paper helps to get more significant and meaningful results and to identify weaknesses before the deployment in real world applications.

\bibliographystyle{springer_short}
\bibliography{IEEEabrv,multihop_testbeds}

\label{lastpage} 
\end{document}